\documentclass[aps,prl,twocolumn,superscriptaddress]{revtex4-1}
\usepackage{graphicx}
\usepackage{natbib}
\usepackage{amssymb} %for Box, qed.
\newcommand{\ex}[1]{\langle #1 \rangle}

\newcommand{\ket}[1]{| #1 \rangle}
\newcommand{\bra}[1]{\langle #1 |}
\newcommand{\rb}[1]{\left( #1 \right)}

\newcommand{\ew}[1]{\langle #1 \rangle}
\newcommand{\beq}{\begin{eqnarray}}
\newcommand{\eeq}{\end{eqnarray}}

\newcommand{\op}[2]{| #1 \rangle \langle #2 |}

\newcommand{\eq}[1]{Eq.~(\ref{#1})}
\newcommand{\fig}[1]{Fig.~\ref{#1}}

\begin{document}

%%%%%%%%%%%%%%%%%%%%%%%%%%%%%%%%%%%%%%%%%%%%%%%%%%%%%%%%%%%%%%%5
\title {Distinguishing quantum and classical transport through nanostructures}
\author{Neill Lambert}
\affiliation{Advanced Science
        Institute,
     The Institute of Physical and Chemical Research (RIKEN), Saitama 351-0198, Japan}
     \author{Clive Emary}
\affiliation{ Institut f\"ur Theoretische Physik,
  Technische Universit\"at Berlin,
  D-10623 Berlin,
  Germany}
\author{Yueh-Nan Chen}
\affiliation{Department of Physics and National Center for
Theoretical Sciences, National Cheng-Kung University, Tainan 701,
Taiwan}
\author{Franco Nori}
\affiliation{Advanced Science
        Institute,
     The Institute of Physical and Chemical Research (RIKEN), Saitama 351-0198, Japan}
\affiliation{Physics Department, University of Michigan, Ann
Arbor, Michigan, 48109, USA}

\begin{abstract}
We consider the question of how to distinguish quantum from
classical transport through nanostructures.  To address this issue
we have derived two inequalities for temporal correlations in
nonequilibrium transport in nanostructures weakly coupled to
leads.  The first inequality concerns local charge measurements
and is of general validity; the second concerns the current flow
through the device and is relevant for double quantum dots.
Violation of either of these inequalities indicates that physics beyond that of a classical Markovian model is occurring in the nanostructure.
\end{abstract}
\maketitle
%%%%%%%%%%%%%%%%%%%%%%%%%%%%%%%%%%%%%%%%%%%%%%%%%%%%%%%%%%%%%%%%

%%%%%%%%%%%%%%%%%%%%%%%%%%%%%%%%%%%%%%%%%%%%%%%%%%%%%%%%%%%%%%%%
%%%%%%%%%%%%%%%          INTRODUCTION          %%%%%%%%%%%%%%%%%
%%%%%%%%%%%%%%%%%%%%%%%%%%%%%%%%%%%%%%%%%%%%%%%%%%%%%%%%%%%%%%%%

Quantum coherence of electrons is the essential ingredient behind
many interesting phenomena in nanostructures (e.g.,~\cite{Brandes, Kouwenhoven}).  % which, it is hoped, may one day form the basis of
%new electronic devices %\cite{staff}
%and even a quantum computer.
%\cite{Loss98}.
%
  Considerable progress has recently been made in the investigation
of coherent effects in nanostructures with both charge and
transport measurements,
(e.g.,~\cite{ Kouwenhoven, Shinkai, *Hayashi03,*Fujetal98,Kiesslich, Petta05sci,PNAS, FujiQPC,*Gustav}).  %Typically, e.g.
%\cite{Fujetal98,Hayashi03}, oscillations (
Typically Rabi oscillations in the current are taken as a
distinctive signature of quantum coherence.  However, since even
classical autonomous rate equations can admit oscillatory
solutions (e.g.,~\cite{Timm,*Omelyanchouk, *Wei}), oscillations by
themselves cannot be considered as a definitive proof of the
existence of quantum coherent dynamics.

In this paper we formulate a set of inequalities that would allow
an experimentalist to exclude the possibility of a classical
description of transport through a nanostructure.  The inspiration
for this comes from the Leggett-Garg inequality \cite{LG1,*LG2},
which has been described as a single-system temporal version of
the famous Bell inequality, %\cite{Bell64,CHSH},
also a topic of interest in nanostructures at the present
(e.g.,~\cite{martinis,*Emary04,*weiBell}). The Leggett-Garg
inequality \cite{LG1,*LG2} can be summarized as follows. Given an
observable $Q(t)$, which is bound above and below by
$|Q(t)|\leq1$, the assumption of: (A1) macroscopic realism and
(A2) non-invasive measurement implies the inequality,
%\beq
%   \ex{Q(t)Q(t+\tau_1)} &+&
%   \ex{Q(t+\tau_1)Q(t+\tau_1+\tau_2)}\nonumber\\ &-&
%   \ex{Q(t)Q(t+\tau_1+\tau_2}\leq 1
%   \label{LG}.
%\eeq
\beq
        \ew{Q(t_1) Q }
        +
        \ew{Q(t_1\!+\!t_2)Q(t_1)}
        -
        \ew{Q(t_1\!+\!t_2)Q}
        \leq 1
    \label{LG}
    ,\ \ \
\eeq
%One can see that a measurement which conforms to (A1) and (A2) cannot violate this bound.
where $Q\equiv Q(t=0)$. %The %main physical
%focus in~\cite{LG1,*LG2} is on
The question of (A1) `realism'~\cite{LG1,*LG2} can be phrased as:
{\em before} we perform the measurement $Q$ on the
system~\cite{Sahelll}, does it have a well defined value? A
classical system does, but a quantum system does not.

In the context of nanostructures weakly coupled to contacts, such
that a generalized master equation description
(e.g.,~\cite{Gurvitz96,*Gurvitz98,Brandes,petri}) is appropriate,
we derive and study two inequalities.
The first concerns correlations between local charge measurements
performed, e.g., by a quantum point contact (QPC)
(e.g.,~\cite{FujiQPC,*Gustav}).  We formulate this inequality in
quite general terms, applicable to a range of nanostructures.  The
charge measurements we consider here are related, in spirit, to
recent work (e.g.,~\cite{korot1,*korot2,*jordan,*vion}) on
violations of the Leggett-Garg inequality, using continuous weak
measurements on closed systems. However, in contrast to their
work, here we are considering a very different situation: ensemble
averages of {\em strong\/} (i.e., projective) non-continuous
measurements on {\em open\/} transport systems. Moreover, our
second inequality explicitly focuses on DQDs, providing an
inequality for the correlation functions of the current flowing
through this widely-studied nanostructure. This second inequality
is of particular relevance to DQD experiments along the lines of
those of Ref.~\cite{Shinkai, *Hayashi03,*Fujetal98,Kiesslich},
where we predict that violations
of both inequalities should occur. %be visible.

%{\tt relation to Korotkov's recent work on Leggett-Garg inequality violations with weak measurement\cite{korotkov}?}.

%%%%%%%%%%%%%%%%%%%%%%%%%%%%%%%%%%%%%%%%%%%%%%%%%%%%%%%%%%%%%%%%
\begin{figure}[]
    \includegraphics[width=\columnwidth]{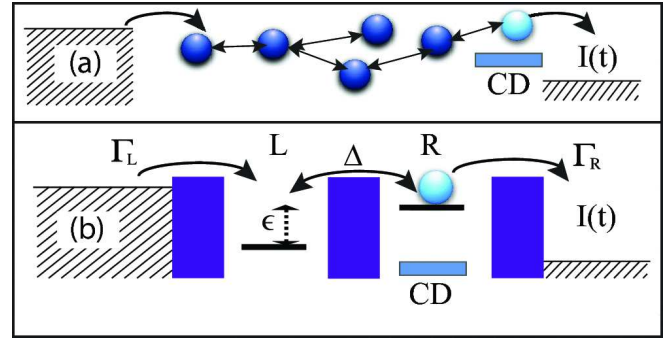}
    \caption{(Color online)
   (a) shows a generic
single-charge nanostructure.  The single-charge can occupy one of
$N$ internal states.  Local charge occupation of one or more sites
is measured using a charge detector (CD) \eq{LQineq}, e.g. a
quantum point contact. (b) shows a schematic of a double quantum
dot, the main example we discuss here. $\Gamma_{L/R}$ describe the
left/right tunnelling rates, $\Delta$ is the coherent tunnelling
amplitude between the left and right dots, and $\epsilon$ is the
energy difference between the left and right occupation states.
Local charge measurements are done on the right ($R$) state for
the charge inequality of \eq{LQineq}, or transport is measured
into the collector reservoir for the current inequality of
\eq{LI}.
        \label{systemsfig}
    }
\end{figure}
%%%%%%%%%%%%%%%%%%%%%%%%%%%%%%%%%%%%%%%%%%%%%%%%%%%%%%%%%%%%%%%%

%%%%%%%%%%%%%%%%%%%%%%%%%%%%%%%%%%%%%%%%%%%%%%%%%%%%%%%%%%%%%%%%
%%%%%%%%%%%%%%%            SYSTEMS             %%%%%%%%%%%%%%%%%
%%%%%%%%%%%%%%%%%%%%%%%%%%%%%%%%%%%%%%%%%%%%%%%%%%%%%%%%%%%%%%%%
{\it Systems.--- } We begin by outlining the class of systems
studied in this work.  We consider nanostructures
(\fig{systemsfig}) weakly coupled to leads  such that transport
proceeds via sequential tunnelling, and we assume a large bias
such that higher-order tunnelling, level-broadening, and
non-Markovian effects can be neglected
\cite{Gurvitz96,*Gurvitz98}. We assume strong Coulomb blockade
such that the system admits at most one excess electron. In these
limits the master equation formalism we apply here, while simple,
has been shown to be very accurate by a variety of
experiments~\cite{ Kouwenhoven, Shinkai, *Hayashi03,*Fujetal98,
Kiesslich, Petta05sci,PNAS, FujiQPC,*Gustav}.   In general,
non-Markovian effects might lead to a violation of these
inequalities, so care must be taken to verify one is in these
limits.
%Examples include networks of quantum
%dots, Cooper pair boxes, and molecules.
%
Our %description of the
 system comprises of $(N+1)$ states: the
``empty'' state, $\ket{0}$, with no excess electron, and states $
\ket{n}$ with a single excess electron in state $n = 1 \ldots N$.
The dynamics is described by the generalized master equation
$\dot{\rho}(t) =\mathcal{W}[\rho(t)]$, where $\rho(t)$ is the
density matrix and the superoperator $\mathcal{W}$ is the
Liouvillian.
Within a quantum-mechanical description, the density matrix
$\rho(t)$ contains coherences, and the Liouvillian has a Lindblad
form $\mathcal{W}=\mathcal{W}_0 + \Sigma$, where the coherent
evolution of the system is given by $\mathcal{W}^\mathrm{qm}_0 =
-i \left[H,\cdot\right]$ with $H$ the internal system Hamiltonian,
$\Sigma$ is the self-energy induced by the contact with reservoirs
(both electronic and otherwise), and throughout we set
$e=\hbar=1$.

The analogous classical description is a rate equation for the
probabilities $P_n(t)$ of finding the system in state $n$ at time
$t$.  This rate equation can be written in the same form as above:
$\dot{\rho} =
  \mathcal{W}^\mathrm{cl}[\rho]$, but now the density matrix $\rho$ only includes diagonal elements
   [the probabilities $P_n(t) = \rho_{nn}(t)$].  This $\rho$ can be represented as a vector such that the Liouvillian  ${\cal W}^\mathrm{cl}$ is a
    classical rate matrix with ${\cal W}^\mathrm{cl}_{ij}>0;i\ne j$ and ${\cal W}^\mathrm{cl}_{ii}=-\sum_{j\ne i} {\cal W}^\mathrm{cl}_{ji}$.
Our general strategy is to explore the behavior allowed by this
classical rate equation and use this to derive our inequalities.

%%%%%%%%%%%%%%%%%%%%%%%%%%%%%%%%%%%%%%%%%%%%%%%%%%%%%%%%%%%%%%%%
%%%%%%%%%%%%%%%       CHARGE INEQUALITY        %%%%%%%%%%%%%%%%%
%%%%%%%%%%%%%%%%%%%%%%%%%%%%%%%%%%%%%%%%%%%%%%%%%%%%%%%%%%%%%%%%

{\it Charge Inequality.--- } The first inequality we derive is for
localized state measurements.  Consider a charge detector which
registers the value $Q_n\geq 0$ when the system is in state $n$
(see Fig.~\ref{systemsfig}). Furthermore, let us designate as
state $N$ the state for which $Q$ has maximum value: $Q_N =
Q_\mathrm{max}$.

{\it (i) Classical regime:} We assume that for a classical system
the charge measurement can be performed non-invasively.  An
initial state, described by a set of probabilities $P_n(0)$, is
fixed and known (actually one only requires knowledge of the
relevant expectation values in this state).  When non-invasively
measuring the charge of a classical Markovian system, we posit
that the following
inequality holds %we posit that for classical Markovian systems
%with noninvasive charge measurements, the following inequality holds
\beq
  |L_Q(t)| \equiv |2\ew{Q(t)Q} - \ew{Q(2t)Q} |
  \leq Q_\mathrm{max} \ew{Q}
  \label{LQineq}
  ,
\eeq where $\ew{Q} = \sum_k P_k(0) Q_k$ is the expectation value
of $Q\equiv Q(t=0)$, % in the initial state,
and $\ew{Q(t)Q}$ is the charge-charge correlation function.  This
inequality holds in two regimes: (i) stationarity, where it
follows from the original Leggett-Garg [Eq.(1)] by defining the
normalized operator $Q= 2Q/Q_\mathrm{max}-1$~\cite{footnote} and
taking the stationary expectation value; and (ii) if only a single
state contributes to the detection process, i.e., $Q_n =
Q_\mathrm{max} \delta_{nN}$, then \eq{LQineq} holds for an
arbitrary initial state (defined by the set of probabilities
$P_k$), and not just the stationary state. This latter can be seen
as follows.  Within both classical and quantum stochastic theory,
the charge-charge correlation function can be written as $
  \ex{Q(t)Q} = Q_\mathrm{max} \Omega_{NN}(t) Q_\mathrm{max} P_N(0)
%  \label{QQPN}
  ,
$ where the ``propagator'' $\Omega_{NN}(t)$ is an element of the
stochastic matrix giving the probability of finding the system in
local charge state $N$ a time $t$ after it is in state $N$.  The
quantity $L_Q$ can thus be written as (for this single state
measurement) $
  L_{Q}(t) =
   Q_\mathrm{max}^2 P_N(0) \left[2\Omega_{NN}(t)-\Omega_{NN}(2t)\right]
  \label{L1}
$.
 If the behavior is classical and Markovian, then the
Chapman-Kolgomorov equation for classical rate equations applies
\cite{petri}, and we can write the propagator with argument $2t$
as a decomposition over intermediate states $
  \Omega_{NN}(2t) = \sum_k \Omega_{Nk}(t)\Omega_{kN}(t)
  \label{PNNdecomp}
$ to obtain $
  L_{Q}(t) =
   Q_\mathrm{max}^2 P_N(0)
   [
     \Omega_{NN}(2-\Omega_{NN})
     -\sum_{k\ne N} \Omega_{Nk}\Omega_{kN}
   ]
$.
Hereafter, we suppress the time argument, $\Omega=\Omega(t)$.
$L_{Q}(t)$ is then {\em maximized} by choosing the propagators
such that the system always ends up in state $N$,  i.e.
$\Omega_{NN}=1$, which gives: $\mathrm{max}\{L_{Q}(t)\} =
Q_\mathrm{max}^2 P_N(0) = Q_\mathrm{max} \ew{Q}$.  The lower bound
is: $\mathrm{min}\{L_{Q}(t)\} = - Q_\mathrm{max} \ew{Q}$.  For
this single state measurement the inequality holds independent of
initial state, as the dynamics are sufficiently constrained by the
Chapman-Kolgomorov equation alone.  %The original Leggett-Garg
%inequality requires us to use a full three-time form \cite{LG1} if
%the initial conditions are non-stationary.
   %In this stationary limit, time-translational invariance
%allows the original Leggett-Garg inequality in \eq{LG} to be
%recast for the above charge measurements, with $t_1=t_2=t$, as:
%$2\ew{Q(t)Q} - \ew{Q(2t)Q}  \leq Q_\mathrm{max}^2$. Since $ \ew{Q}
%\leq Q_\mathrm{max}$, the inequality of \eq{LQineq} presents a
%tighter upper bound than the original Leggett Garg inequality
%\textbf{{\em However, as mentioned earlier, this bound
%(\eq{LQineq}) can also be derived from the Leggett Garg inequality
%if one chooses a measurement whose eigenvalues span $\{-1,+1\}$,
%e.g. $\sigma_z$ for a qubit. However our derivation above is
%important to found a bound for invasive measurements, like the
%current.
We first illustrate the charge inequality violation with
an example, before continuing to derive the inequality for current
measurements.

%%%%%%%%%%%%%%%%%%%%%%%%%%%%%%%%%%%%%%%%%%%%%%%%%%%%%%%%%%%%%%%%
%%%%%%%%%%%%%%%         DOUBLE QUANTUM DOT     %%%%%%%%%%%%%%%%%
%%%%%%%%%%%%%%%%%%%%%%%%%%%%%%%%%%%%%%%%%%%%%%%%%%%%%%%%%%%%%%%%

{\it (ii) Quantum regime:} The transport DQD consists of a dot
$L$, connected to the emitter, and dot $R$, connected to the
collector (see Fig.~\ref{systemsfig}).
Assuming weak coupling, large bias, and Coulomb Blockade, the
basis of electron states is $\{\ket{0}, \ket{L}, \ket{R}\}$.  Its
Hamiltonian becomes \beq
  H=\epsilon (\op{L}{L} -\op{R}{R})
  + \Delta(\op{L}{R}+ \op{R}{L})
  ,
\eeq
with $\epsilon$ the level splitting, and $\Delta$ the
coherent tunnelling amplitude betweens the dots, and with
self-energy \beq
  \Sigma [\rho] &=&
  -\frac{1}{2}\sum_{\alpha=L,R} \Gamma_\alpha
  \left[s_\alpha s_\alpha^\dagger \rho
    -
    2s_\alpha^\dagger \rho(t)s_\alpha + \rho s_\alpha s_\alpha^\dagger\right]
    \nonumber
    ,
\eeq where $s_L=\ket{0}\bra{L}$, $s_R=\ket{R}\bra{0}$, and
$\Gamma_L$ and $\Gamma_R$ are the left/right tunnelling rates
(throughout we set $e=\hbar=1$).  The influence of phonons can
also be included in $\Sigma$ in the standard way
\cite{Brandes,footnote2}.
%, and the affect is discussed in the figure captions.
%
The corresponding classical Liouvillian is a $3\times3$ matrix with elements ${\cal W}^\mathrm{cl}_{\alpha\beta}$; $\alpha,\beta = 0,L,R$.
For illustrative purposes, we consider a charge measurement in
which the detector only registers when there is an electron in the
righthand QD: $Q = \op{R}{R}$ for which $Q_{\mathrm{max}}=1$.
The correlation functions are then calculated from $\ex{Q(t)Q} =
\mathrm{Tr} \left[Q e^{\mathcal{W} t} Q \rho_0 \right]$, with
$\rho_0$ the stationary density matrix of the system.% obtained
%from the null-space of the Liouvillian.

\begin{figure}[]
    \includegraphics[width=\columnwidth]{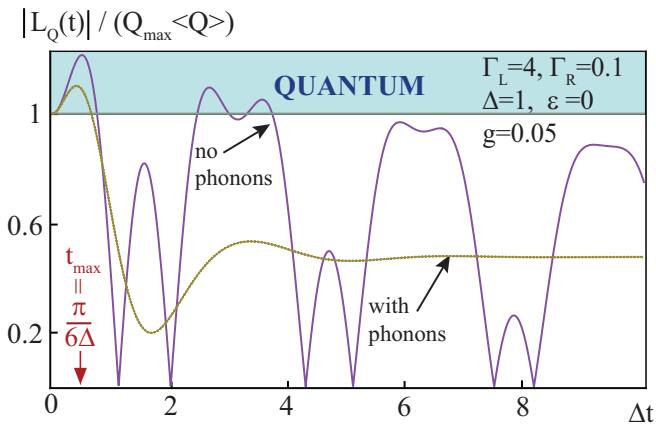}
    \caption{
      (Color online)  The
charge inequality $|L_Q(t)|/(Q_{\mathrm{max}}\ex{Q})$ \eq{LQineq}
for QPC charge measurements on the $R$ state of a double quantum
dot, as a function of dimensionless time $\Delta t$. The
parameters are shown on the figure, and we have set
$Q_{\mathrm{max}}=1$. We chose a dimensionless coupling constant
$g=0.05$ for the bulk phonons \cite{Brandes}. The solid line
represents the case of no phonons, and the dashed line includes a
phonon bath at temperature $T=10\hbar \Delta/k_B$, to illustrate
how an invasive environment masks the violation. The colored
region marks the area of violation of the inequality in
\eq{LQineq}.
        \label{LQ_fig}
    }
\end{figure}

In \fig{LQ_fig} we plot $|L_Q(t)|/(Q_\mathrm{max} \ew{Q})$ as a
function of time for a DQD.
The behavior is oscillatory, but also damped due to coupling to
both the collector and the phonon bath.
The shaded region ($>1$) indicates where $|L_Q(t)|$ violates the
inequality of~(\ref{LQineq}).  The most prominent violation occurs
at the maximum closest to $t=0$.  For these parameters then, no
classical Markov description of the system is possible and here,
it is quantum oscillations between $L$ and $R$ states that are
responsible for the violation.
As we discuss later, the degree of violation can be increased by
decreasing $\Gamma_R$, which permits the electron to spend a
longer time in the DQD.  In the limit $\Gamma_R\rightarrow 0$ and
with $\Gamma_L\rightarrow \infty$, such that the empty state may
be eliminated,  we find the analytic form $L_Q(t)/(Q_\mathrm{max}
\ew{Q})= \left[\cos(2\Delta t)+\sin^2(2\Delta t)\right]$, for
$\epsilon =0 $. The coherent tunnelling $\Delta$ defines the time
when the violation is maximum, $t_{\mathrm{max}}=\pi/6\Delta$,
such that
$L_Q(t_{\mathrm{max}})/(Q_\mathrm{max} \ew{Q}) = \frac{5}{4}$.  %  With
%$\ew{Q}=????$, the maximum degree to which the inequality is
%violated is therefore $L_Q(t_{\mathrm{max}})/Q_\mathrm{max} \ew{Q}
%= ?????$. {\tt NL: need this from your calc!}.
%
  The time $t_{\mathrm{max}}$ is in agreement with that observed for the Leggett-Garg inequality ~\cite{LG1} for a single free qubit with
level coupling $\Delta/2$.
The effects of a phonon bath are also apparent in \fig{LQ_fig},
where we have used reasonable bath parameters~\cite{Brandes}.
Although the oscillations of $L_Q(t)$ are damped, the first and
most significant maximum remains.

%%%%%%%%%%%%%%%%%%%%%%%%%%%%%%%%%%%%%%%%%%%%%%%%%%%%%%%%%%%%%%%%
%%%%%%%%%%%%%%%%%%%%%%%%%%%%%%%%%%%%%%%%%%%%%%%%%%%%%%%%%%%%%%%%

%%%%%%%%%%%%%%%%%%%%%%%%%%%%%%%%%%%%%%%%%%%%%%%%%%%%%%%%%%%%%%%%
\begin{figure}[]
    \includegraphics[width=\columnwidth]{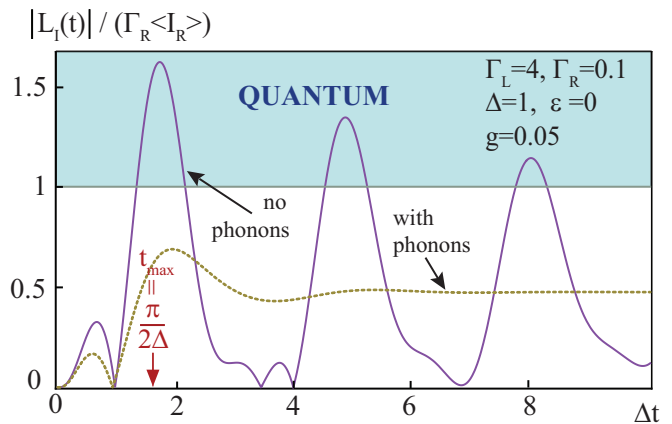}
    \caption{
(Color online) The current inequality
$|L_I(t)|/(\Gamma_R\ex{I_R})$ \eq{LI} for current measurements on
a double quantum dot, as a function of dimensionless time $\Delta
t$.   The solid line represents the case of no phonons, and the
dashed line includes a phonon bath at $T=10\hbar \Delta/k_B$. The
colored region marks the area of violation of the inequality in
\eq{LI}.  The violation of this inequality is more pronounced than
that of the charge inequality.  The effect, however, is more
sensitive to the phonon bath.
        \label{DQDcurrentfig}
    }
\end{figure}
%%%%%%%%%%%%%%%%%%%%%%%%%%%%%%%%%%%%%%%%%%%%%%%%%%%%%%%%%%%%%%%%

%%%%%%%%%%%%%%%%%%%%%%%%%%%%%%%%%%%%%%%%%%%%%%%%%%%%%%%%%%%%%%%%
%%%%%%%%%%%%%%%         CURRENT INEQUALITY     %%%%%%%%%%%%%%%%%
%%%%%%%%%%%%%%%%%%%%%%%%%%%%%%%%%%%%%%%%%%%%%%%%%%%%%%%%%%%%%%%%
{\it Current Inequality.--- } Our second inequality concerns the
current $I(t)$ flowing through the transport DQD: \beq
 |L_I(t)| \equiv
 |2\ew{I(t)I} - \ew{I(2t)I} | \le \Gamma_R \ew{I},
 \label{LI}
\eeq where $\Gamma_R$ is the coupling to the collector, $I\equiv
I(t=0)$ and $\ew{I}$ is the average current of the initial state.
Although this second inequality resembles the first one, in
\eq{LQineq} ($\Gamma_R$ is the maximum instantaneous collector
current), its derivation and significance are somewhat different.
This is because, in the master equation approach, the current
operator translates into a ``jump'' super-operator and \eq{LI}
thus represents an inequality concerning transitions in the
system, and not static properties such as the charge under the
noninvasive measurements of \eq{LQineq}.
For the DQD model in the infinite bias limit, the current
super-operator acts as $
    {\cal J} [\rho] = \Gamma_R \ket{0}\bra{R}\rho
    \ket{R}\bra{0}
$,
such that the average current is $\ew{I} = \mathrm{Tr}\left\{{\cal J} \rho\right\}$ and the correlation function of interest is obtained as
$
  \ew{I(t)I}
  =
  \mathrm{Tr}\left\{
    {\cal J}e^{\mathcal{W}t}{\cal J}
  \rho_0\right\}
$, where again the stationary distribution is chosen as the
initial state.
In these terms, \eq{LI} can be written as $
  L_I(t)=
  \mathrm{Tr}
  \left\{
     \mathcal{J}
    \rb{
      2 e^{\mathcal{W}t}-e^{2 \mathcal{W}t}
    }
    \mathcal{J}
  \rho_0\right\}.
$
  In the classical description of the DQD, ${\cal J} $ is the $3\times3$ matrix with elements ${\cal J}_{\alpha\beta} = \Gamma_R \delta_{\alpha,0}\delta_{\beta,R}$.  Thus using Chapman-Kolgomorov again, we have
$
  L_I(t)
  =
  \Gamma_R^2 P_R(0)
  \rb{
    \Omega_{R0}\rb{2-\Omega_{00}-\Omega_{RR}}
    - \Omega_{RL}\Omega_{L0}
  }.
  \nonumber
%  \label{Ltil3LS}
$
For a general Markov stochastic matrix, $\Omega$, the maximum of $L_I$ is $ 2\Gamma_R^2 P_R(0) $. However, the rate equation form $\Omega(t) = e^{{\cal W} t}$ furnishes us with a further constraint.
Maximizing $L_I(t)$ with respect to time, from $ \dot{L}_I=0$ and
$\dot{\Omega} = {\cal W}\Omega$, we find that the maximum of $L_I$
occurs when $\Omega_{00} +\Omega_{RR}=1$ and $\Omega_{R0}=1$,
giving $\mathrm{max}\{L_I\} = \Gamma_R^2 P_R(0) = \Gamma_R
\ew{I}$. This result relies on the geometry of the DQD, and in
particular the form of the jump operator and the absence of direct
tunnelling from emitter to dot $R$, i.e.  ${\cal W}_{R0}=0$.
%\hfill\ensuremath{\Box}

Figure \ref{DQDcurrentfig} illustrates the violation of the current inequality \eq{LI} for the DQD.
As with the charge measurement, the quantity $L_I(t)$ is oscillatory and violates the respective inequality with
the strongest violation occurring at the first maximum, which is here at a time $t_{\mathrm{max}}= \pi/(2\Delta)$.
The degree to which this current inequality is violated is of greater magnitude than that for the charge measurement.
Again, in the limit $\Gamma_L \rightarrow \infty$, one can
eliminate the empty state and find an analytical form. In
addition, the $\Gamma_R \rightarrow 0$ limit gives
$L_I(t)/(\Gamma_R\ew{I})=-2\sin^2(\Delta t)\cos(2\Delta t)$.  Thus
the violation has a maximum of
$L_I(t_{\mathrm{max}})/(\Gamma_R\ew{I})=2$.

Under the above assumptions, the three-state classical DQD
Liouvillian cannot produce a violation.  However, if these
assumptions are relaxed (e.g., allowing ${\cal W}_{R0}\neq 0$), a
 small violation (on the order of $0.003\% \ \Gamma_R \ex{I}$) of the inequality can be observed in extreme
parameter regimes. This is in contrast to the $L_Q$ inequality,
where no further constraints are required of the Liouvillian. This
difference reflects the fact that here the current measurement is
essentially a destructive measurement of the state of the
nanostructure.  An infinitesimal time interval after a positive
current measurement is obtained, the electron has left the system,
leaving it in the `empty' state.  This behavior is implicit in the
jump super-operator form of the current measurement.%% and is akin
%%to the
%%destructive measurement of a photon.

    % in quantum optics. %In line
%with this destructive nature of the measurement, $L_I(t)$ tends to
%zero for $t\rightarrow 0$. {\tt I don't understand this}.

%%%%%%%%%%%%%%%%%%%%%%%%%%%%%%%%%%%%%%%%%%%%%%%%%%%%%%%%%%%%%%%%%
\begin{figure}[]
    \includegraphics[width=\columnwidth]{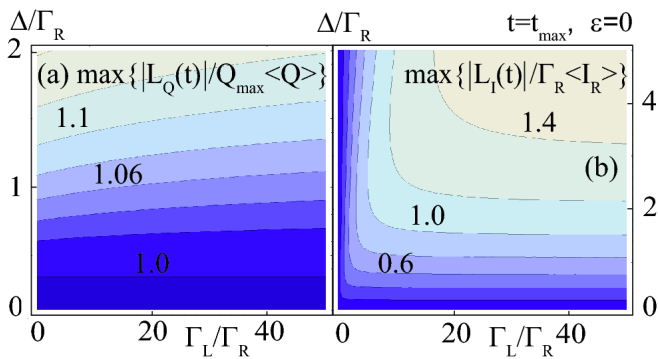}
    \caption{(Color online) Contour plot of (a) $\mathrm{max}\{L_Q(t)/Q_{\mathrm{max}}\ex{Q}\}$ and
(b) $\mathrm{max}\{L_I(t)/\Gamma_R\ex{I_R}\}$ as a function of the
tunnelling amplitude $\Delta/\Gamma_R$ and relative rates
$\Gamma_L/\Gamma_R$.  As discussed in the text, violation requires
$ \Delta \gtrsim 0.3\Gamma_R$ in (a) and, in (b), $ \Delta \gtrsim
1.5\Gamma_R$, $\Gamma_L/\Gamma_R \gg 1$. The phonon bath is
excluded. \label{MAXfig} }
\end{figure}
%%%%%%%%%%%%%%%%%%%%%%%%%%%%%%%%%%%%%%%%%%%%%%%%%%%%%%%%%%%%%%%%%

Figure \ref{MAXfig} shows how the maximum degree of violation of
both inequalities depends on the parameters of the DQD with no
phonons.  A violation of the current inequality~\eq{LI} requires
$\Delta \gtrsim 1.5\Gamma_R$, $\Gamma_L \gg \Gamma_R$, and small
detuning ($\epsilon<\Delta$).  The violation of the charge
inequality~\eq{LQineq} is more resilient, and always occurs unless
there is strong over-damping from the reservoir $\Gamma_R \gg
\Delta$.

Finally, %we point out that the maximum violation for both
%inequalities is found to occur on the time scale of $1/\Delta$.
%Thus,
we note that in practice one needs to measure the correlation
functions in \eq{LQineq} or \eq{LI} on very short time scales
(e.g.,~\cite{FujiQPC,*Gustav}). Alternatively, one can obtain
either correlator from the inverse Fourier transform of the
appropriate %high-frequency
noise power function. In the transport
case (\eq{LI}), one must consider contributions from both
particle- (as considered here) and displacement-currents. In
principle, one can either choose appropriate gate/junction
capacitances to neglect the displacement current contribution, or
include them in the definition of \eq{LI}, and its subsequent
maximization.

%%%%%%%%%%%%%%%%%%%%%%%%%%%%%%%%%%%%%%%%%%%%%%%%%%%%%%%%%%%%%%%%
%%%%%%%%%%%%%%%           CONCLUSIONS          %%%%%%%%%%%%%%%%%
%%%%%%%%%%%%%%%%%%%%%%%%%%%%%%%%%%%%%%%%%%%%%%%%%%%%%%%%%%%%%%%%

{\it Conclusions.--- }  In summary, we have derived two
inequalities for non-equilibrium transport in nanostructures: one
concerning local charge measurements and the other for current
flow through the device. The first is of general validity; the
second of relevance to the usual DQD geometry found in numerous
experiments.
Violation of either of these inequalities indicates that physics
beyond that of a classical Markovian model is occurring in the
nanostructure.  This may be taken as evidence for quantum
oscillations of the electron within the device; or it may indicate
a non-Markovian interaction with previously unappreciated degrees
of freedom.
Finally, we point out that these ideas can be expanded in a number
of different directions: to other types of measurements; to
inequalities with different time-dependencies, as in the original
Leggett-Garg work \cite{LG1,*LG2}; and to different physical
situations for which master equations are appropriate, such as
atom-field interactions in quantum optics.  This work can also be
applied to networks of quantum dots, Cooper pair boxes, and
molecules.

\acknowledgments

%We thank S. Ashhab, A. Kofman, and T. Brandes for helpful
%discussions.
NL is supported by RIKEN's FPR Program.  YNC is supported
partially by the NSC, Taiwan.%, under the grant no.
%98-2112-M-006-002-MY3.
%
 CE is supported by the WE Heraeus Foundation and by DFG grant BR
1528/5-1.
FN acknowledges partial support from the NSA, LPS, ARO, NSF grant
No.~0726909, JSPS-RFBR contract No.~09-02-92114, MEXT Kakenhi on
Quantum Cybernetics,  and FIRST (Funding Program for Innovative
R\&D on S\&T).

\vspace*{-0.1in} \vspace*{-0.1in}

\bibliography{bibliography}

\begin{thebibliography}{10}
\bibitem{footnote2}%
  \BibitemOpen
\bibfield{journal}{%
    }%
  \bibinfo {journal} {The rates for the electron-phonon interaction are,
  $\gamma_{\pm}=[-\epsilon (T/\Delta^2) \text{Coth}[\beta \Delta /2]\pm
  T/\Delta ]\pi /2 J[\Delta ]$, $\gamma=(T^2/\Delta^2)\text{ }\text{Coth}[\beta
  \Delta /2] 2 \pi J[\Delta ]$, where $\Delta =\sqrt{\epsilon^2+4 T^2}$,
  $J[\Delta]\text{:=}g \Delta e^{[-\Delta/\Omega_c]}$, $g$ is the dimensionless
  coupling strength, $\beta=(\text{k}_bT)^{-1}$ and $\Omega_c$ is an
  ultraviolet cutoff.}%
  \bibAnnoteFile{Stop}{footnote2}%
\end{thebibliography}

\end{document}